# Radiation Reaction Effects in Cascade Scattering of Intense, Tightly Focused Laser Pulses by Relativistic Electrons

A. Zhidkov[1,2], S. Masuda[1,2], S.S. Bulanov[3], T. Hosokai[1,2], J. Koga[4], R. Kodama[1,5]

[1] *Photon Pioneers Center, Osaka University, 2-1, Yamadaoka, Suita, Osaka 565-0871, Japan*

[2] *Japan Science and Technology Agency, CREST, 2-1, Yamadaoka, Suita, Osaka 565-0871, Japan*

[3] *University of California, Berkeley, California 94720, USA*

[4] *Kansai Photon Science Institute, JAEA, Kizugawa, Kyoto 619-0215, Japan*

[5] *Graduate School of Engineering; Institute of Laser Engineering, Osaka University, 2-1, Yamadaoka, Suita, Osaka 565-0871, Japan*

## Abstract

Non-linear cascade scattering of intense, tightly focused laser pulses by relativistic electrons is studied numerically in the classical approximation including the radiation damping for the quantum parameter $\hbar\Omega_{\text{x-ray}}/\varepsilon < 1$ and an arbitrary radiation parameter $\chi$. The electron energy loss, along with its side scattering by the ponderomotive force, makes the scattering in the vicinity of high laser field nearly impossible at high electron energies. The use of a second, co-propagating laser pulse as a booster is shown to solve this problem.

The scattering of intense laser light by relativistic electrons, especially in cascade regimes when an electron scatters more than one x-ray photon, is one of the most important mechanisms of radiation from plasma produced by intense femtosecond laser pulses [1]. Presently, the laser intensities are approaching to a threshold where the radiation becomes dominant [2] and, therefore, the matter of detailed study. So far, ab initio simulations of the radiation even from a classical relativistic plasma standpoint, requiring extremely high spatial and temporal resolution, are impossible. Understanding of the radiation processes in relativistic systems in high laser fields is a pressing issue to derive proper empirical approaches for calculation of the radiation damping and spectra of such plasma. Non-linear and quantum radiation effects in the interaction of relativistic electron beams and intense laser pulses occur at relatively smaller, accessible laser intensities. Their experimental and numerical study could result in developing the necessary radiation models.

In contrast to the cyclotron radiation [3], the laser light scattering process is always a free-free electron transition and may run differently depending on the field and electron energies. At low laser pulse intensities and high electron energies there is the well-known Compton scattering determined by the only quantum parameter: $\nu = 2\gamma_0 \hbar \omega_0 / mc^2$, $\gamma_0$ is the electron relativistic factor and $\omega_0$ is the frequency of a laser photon (see detail in Ref. [4]). If the parameter $\nu$ exceeds the unity the maximal energy of photons must be less than $4\gamma^2 \omega_0$ and is emitted in an angle $\pi - \delta\theta$. The scattering cross-section decreases with further $\gamma_0$ increase [4]. At high intensities the 'quantum' scattering is usually characterized by a parameter: $\chi = (\hbar \omega_0 \gamma / mc^2)\sqrt{(\tilde{\vec{E}} + \vec{v} \times \tilde{\vec{B}}/c)^2 - (\vec{v} \cdot \tilde{\vec{E}}/c)^2}$ [5,6], where

$\tilde{E}(\tilde{B}) = e\vec{E}(\vec{B})/mc\omega = \vec{a}_0$. For $a_0 \gg 1$: $\chi \sim \hbar\omega_0\gamma a_0/mc^2$. At $\chi \gg 1$ the scattering probability, $P$, differs from the classical as $\sim P_{classical}(\chi)\chi^{-1/3}$ [5]. However at $a_0 \ll 1$, this parameter differs from the quantum parameters, $v$, for the Compton scattering [4]. Moreover, the parameter $\chi$ amazingly fits with the part of classical radiation force responsible for the scattering [7]. If the parameter $\chi\tau$, where $\tau$ is the interaction time, is large, the approximation of a 'fixed' electron trajectory as in Ref. [8] is not valid anymore; the radiation damping becomes essential. Depending on the pulse duration $\tau$, it may be a single act resulting in the electron retardation or a cascade of scattering with the same result. The simplest estimation of the number of scattered photons $n_{ph} = \pi(mc^2/e)^2(I\tau/\hbar\omega_0)$, $I$ is the laser pulse intensity, gives $n_{ph} \sim 2$-$3$ or the cascade already at $\sim 1$J energy laser pulse focused in $\sim 10$ μm focus spot.

In a strong laser field when $a_0 \geq 1$ harmonics appear and the quantum parameter is different from the classical. Its definition is not simple. In the 'fixed' trajectory classical approximation the spectrum is flat till the energy is less than the critical energy [8]:

$$\hbar\omega_c = \frac{3a_0^3\hbar\omega_0}{4\sqrt{2}}\left[\frac{\gamma^2(1+v/c)^2}{1+a_0^2/2}+1\right]$$

At high frequencies the spectrum exponentially decreases $\sim\exp(-2\omega/\omega_c)$. The simplest definition of the quantum condition might be $\hbar\omega_c \gg mc^2(\gamma-1)$ (or for large $a_0$ and $\gamma$ $a_0 \gg (\sqrt{2}mc^2/6\hbar\omega_0)/\gamma$ [for XFEL this condition achieves $\sim 10^4$ times lower $a_0$] when the most of radiation energy should be emitted with an essential recoil: parametrically at large $a_0$ this condition coincides with $\chi \gg 1$. However, the radiation damping may change the spectra and result in softer quantum conditions.

The interaction of an electron with an intense laser pulse is not always a stationary process that can be characterized by a cross-section or a probability. For example, a typical quantum consideration of the problem [5,6,9,10] assumes that the electron instantly appears in a strong plane wave propagating contrary to the wave with momentum $p_0$. This approach is built on a relatively simple basis of Volkov's functions, $\psi_p^V = Ae^{iS/\hbar}$, where $A$ is an amplitude, $S$ is a classical action [4,5]. The calculated probability naturally includes the radiation reaction. However, in a strong field an electron will lose its energy rapidly, then, the electron will be trapped and move along the wave. Therefore equations for the scattering matrix

$$\partial C_p / \partial t = -\int dp' \psi_p^V e\hat{\alpha}\hat{A}_{Rad} \bar{\psi}_{p'}^V C_{p'}, \quad C_p(t=-\infty) = \delta_{pp_0},$$ where $A_{Rad}$ is the vacuum radiation potential [4], has no small parameter; their solution $C_p(t \gg \tau)$ restricted by the pulse duration may have a quite broad distribution over momentum $p$ making the exact solution not practical in the cascade regime. Moreover, the ponderomotive scattering is missing in this approach. The classical consideration including the radiation damping may help to understand a physical model necessary for an advanced quantum consideration.

In this letter, the scattering of intense focused laser light and its effects on the electron motion is analyzed in the framework of the classical approximation [11] including the radiation damping and ponderomotive effects.

A 6-th order Runge-Kutta method [12] is used to solve equations of motion for an electron written in the form:

$$\frac{d\vec{p}}{dt} = -e[\vec{E} + \vec{p} \times \vec{B}/(mc\gamma)] - \vec{f}_{RD}; \gamma = \sqrt{1 + \vec{p}^2/(mc)^2}$$

$$\frac{d\vec{r}}{dt} = \vec{p}/(m\gamma)$$

where $f_{RD}$ is the radiation forced in the form [4] with the main part responsible for the light scattering as

$$\vec{f}_{scatt} = -\frac{2e^4}{3m^2c^5}\gamma^2\vec{v}\left\{\left(\vec{E} + \frac{1}{c}\vec{v}\times\vec{H}\right)^2 - \frac{1}{c^2}(\vec{v}\cdot\vec{E})^2\right\}$$

with **E** and **B** the electric and magnetic fields, respectively, and **v** is the electron velocity. To include correctly the ponderomotive force, we use the parabolic approximation for the focused laser field [13]. The transverse components have the following forms

$$E_\perp(r,t) = \exp(-(z/c-t)^2/\tau^2)\sum\left[\operatorname{Re} E^0_{l,m} H_l\left(\frac{\sqrt{2}x}{w(z)}\right) H_m\left(\frac{\sqrt{2}y}{w(z)}\right)\frac{w_0}{w(z)} \times \right.$$
$$\exp(-(x^2+y^2)/w(z)^2 + i(kz - \omega t - (l+m+1)\varphi(z) + k(x^2+y^2)/2R(z))),$$

where for the transverse component of magnetic field equals $\vec{B}_\perp = \pm\vec{E}_\perp$ depending on the propagation direction, $w(z) = w_0\sqrt{1+\frac{z^2}{L_R^2}}$, $R(z) = z\left(1+\frac{L_R^2}{z^2}\right)$, $\varphi(z) = \tan^{-1}(z/L_R)$, and $L_R = \pi w_0/\lambda$ is the Rayleigh length, $w_0 = \pi f\lambda/2D$ [$f$, $D$, and $\lambda$ are the focal length, beam diameter, and the pulse wavelength], $H_l$ is Hermite polynomial. The longitudinal components satisfy equations: div **E**=0 and div**B**=0. Particularly for the pair $\{E_x, B_y\}$:

$E_z = -x[Q(z)/L_R]E_x$, $B_z = -y[Q(z)/L_R]E_x$, and $Q(z) = -i\frac{w_0^2}{w(z)^2} + \frac{L_R}{R(z)}$. Calculations are performed for $l=m=0$ and $w_0=5$ μm. To calculate the spectrum (for the intensity), we exploit the well-known equation for the Lienard-Wiechert potential [7,8]

$$\frac{d^2 I(\omega_{x-ray}, \vec{\Omega})}{d\omega_{x-ray} d\vec{\Omega}} = \frac{e^2}{\pi^2 m^2 c^3} \left| \sum [\vec{n} \times (\vec{n} \times \vec{v}_j)] e^{i\omega_{x-ray}[t - \vec{n}\vec{r}_j/c]} \frac{\sin(0.5\omega_{x-ray}\Delta t(1 - \vec{n}\vec{v}_j/c))}{(1 - \vec{n}\vec{v}_j/c)} \right|^2.$$

Where $r_j$, $v_j$ are the electron coordinate and velocity, respectively, at $t=t_j$, and $\Delta t$ is the integration time step ($\sim 10^{-5} \omega_0^{-1}$).

In Fig.1 the typical spectra of backward scattering ($\theta=0$) are presented for $a_0=1$ for Gaussian pulses 20, 90, and 200 fs with $a_0=1$ for a head-on collision with $\gamma=200$. These spectra expectedly demonstrate the proportional increase of intensity of the scattered x-rays with the pulse duration. However, even in this weakly non-linear case, the spectra depend on the pulse duration. For the long laser pulse the spectrum is very similar to that which follow from a 'fixed' trajectory approximation as in Ref. [8]. For shorter pulses, a certain red shift is clearly seen in the spectrum. We attribute this effect to the stronger ponderomotive force at the entrance for the laser pulse resulting in lower energy of the electron in the vicinity of stronger radiation. The backward spectra for the high intensity laser pulses $a_0=200$ are presented in Fig. 2a,b also for head-on collisions with and without the radiation damping. There is clear dependency on the pulse duration; not only the number of scattered photons increases but there is also clear difference in the spectral distribution. The radiation damping (RD) drastically changes the resulting spectra. The intensity decreases by an order of magnitude when RD is included. The evolution of electron momentum for this case is shown in the inserts. One can see that the radiation in the vicinity of the highest field occurs at much lower electron energy than its initial.

Grey squares exhibit the spectral area where the recoil effect is very strong. Even without RD the integral energy in the quantum is essentially less than that from the lower frequencies emitted without a serious recoil effect. With RD the scattered photons lie far from the quantum area making the quantum contribution softer and, therefore, making the classical approach more accurate. In Fig.3 the parameter $\chi$ is calculated for a relatively high electron energy $\varepsilon$~1 GeV and the strong field with and without RD. One can see that the parameter c reaches ~1 in the absence of RD and a part of the spectra could require the quantum approach. However, RD results in much lower $\chi$ by almost an order of magnitude making the classical calculation quite correct. The spectrum for $\varepsilon$~ 1GeV and different pulse duration are given in Fig. 4 for $a_0$=100 including RD. The difference in the spectrum cannot be expressed in simple probability approaches: while the non-linear part does not change much, the linear part of the spectrum changes dramatically. The increase of $a_0$ up to 500 amazingly results in a smaller number of scattering photons because the main interaction zone occurs at lower pulse intensities even though parameter $\chi$ would reaches ~10 without RD.

The ponederomotive side scattering is an important effect that cannot be easer incorporated in the quantum approach. In Fig. 5 one can see the trajectory of an electron scattered by an intense laser pulse. The calculation is performed for $w_0$=5 μm, however a small shift of the electron 2 μm in the transverse direction results in a dramatic change of its trajectory and spectrum. The most dense electron beams are generated by the laser wake field acceleration [14]. However even those beams have a typical diameter ~10 μm. This means the calculation of the scattered spectrum without the side scattering of the electron during the interaction will result in an incorrect spectrum. We anticipate that the classical calculations may provide reasonable spectral data even for very high energy electron beams. In Fig. 6 the backward spectrum are given for an electron with initial energy $\varepsilon$~8 GeV. One can see the 'quantum' area indicated by the grey color contains relatively low energy and the classical approximation is suitable for electron motion calculations.

The results of calculation show that an efficient Compton scattering in high fields is nearly impossible due to the classical radiation damping or/and the side scattering by the ponderomotive force. A possible solution of the problem is by using two laser pulses. In conventional accelerators the radiation losses are compensated by extra acceleration of electron beams [3]. An electron can be accelerated in vacuum by a co-propagating laser pulse [15]. This acceleration is cycled with the period increasing with $\gamma$ and $a_0$. A proper choice may make the co-propagating laser pulse the electron booster and compensate the energy losses for the radiation during the scattering of the contra-propagating laser pulse.

In Fig. 7, a sample of the evolution of the electron longitudinal momentum in the two pulse scheme is presented. Both pulses have the same intensity $a_0$=100 and duration $\tau$=10 fs, and focused in $w_0$=5 μm. One can see that the radiation damping is compensated by the co-propagating pulse acceleration and the Compton scattering in the vicinity of the maximal laser intensity occurs at the high electron energy. The practical realization of this method is difficult; the applicability of the classical approach for that case requires special analysis.

In conclusion, we have analyzed the effect of the radiation damping in the scattering of intense laser light by relativistic electrons. Nowadays, the classical method is only one, which allows self-consistent analysis of the interaction of relativistic particles with an intense laser field including the ponderomotive side scattering and electron vacuum acceleration. It has been shown that the classical approach is valid at much higher laser intensities and particle energies due to the effect of the radiation damping resulting in much lower frequencies of emitted x-rays. The parameter $\chi$ remains much less than unity due to the radiation damping even for GeV level electrons in the laser intensities $\sim 10^{23}$-$10^{24}$ W/cm$^2$. In the case of plasma, produced by intense short laser pulses, the 'quantum' condition is even softer because plasma electrons acquire their energy from the laser pulses starting from the 'classical' regime.

It has been shown that an efficient experimental realization of the scattering by relativistic electrons is almost impossible due to the radiation damping and ponderomotive side-scattering. Even high energy electrons lose their energy far before the maximum of the laser field. However, the 'two pulse scheme' with a co-propagating laser pulse as an electron booster may solve the problem of efficient non-linear Compton scattering. The ponderomotive acceleration of electrons with the energy increase $\Delta\varepsilon \sim mc^2 \gamma a_0^2/2$ [7,14] can overcome the radiation damping.

**References**

1. G. Mourou, T. Tajima, and S. V. Bulanov, Rev. Mod. Phys. **78**, 309 (2006); M. Marklund and P. Shukla, *ibid.* **78**, 591 (2006);Y. I. Salamin, S. X. Hu, K. Z. Hatsagortsyan, and C. H. Keitel, Phys. Rep. **427**, 41 (2006); A. Di Piazza, C. Muller, K. Z.Hatsagortsyan, and C. H. Keitel, Rev. Mod. Phys. **84**, 1177 (2012).

2. J. Colgan , J. Abdallah, A.Y. Faenov ,et al., Phys. Rev. Letters **110**, 125001(2013); A. Zhidkov, J. Koga, A. Sasaki, and M. Uesaka, Phys. Rev. Lett. **88**, 185002 (2002); S. V. Bulanov, T. Zh. Esirkepov, J. K. Koga, T. Tajima, Plasma Phys. Rep. **30**, 196 (2004), T. Nakamura, J. K. Koga, T. Zh. Esirkepov et al., Phys. Rev. Lett. **108**, 195001 (2012); C. P. Ridgers, C. S. Brady, R. Duclous, et al., Phys. Rev. Lett.**108**, 165006 (2012); C. S. Brady, C. P. Ridgers, T. D. Arber, et al., *ibid.* **109**, 245006 (2012); F.V. Hartemann, LLNL-PROC-409586 (2008)

3. I.M. Ternov, 'Synchrotron radiation', *UFN*, **165**:4 (1995), 429–456

4. V.B. Berestetskii, E.M. Lifshitz, L.P. Pitaevskii, Quantum electrodynamics, V.4 (Butterworth-Neinemann, 1982

5. A. I. Nikishov and V. I. Ritus, Sov. Phys. JETP **19**, 529 (1964);**19**, 1191 (1964); I. Goldman, Sov. Phys. JETP **19**, 954 (1964)

6. N. B. Narozhny, S. S. Bulanov, V. D. Mur, and V. S. Popov, Phys. Lett. A **330**, 1 (2004); S. S. Bulanov, V. D. Mur, N. B. Narozhny, and V. S. Popov, JETP Lett. **80**, 382 (2004); JETP **129**, 14 (2006); S. S. Bulanov, A. M. Fedotov, F. Pegoraro, JETP Lett. **80**, 734 (2004); Phys. Rev. E **71**, 016404 (2005); I. V. Sokolov, J. A. Nees, V. P. Yanovsky, N. M. Naumova, G. A. Mourou, Phys. Rev. E **81**, 036412 (2010); I.

V. Sokolov, N. M. Naumova, J. A. Nees, and G. A. Mourou, Phys. Rev. Lett. **105**, 195005 (2010); S. S. Bulanov, C. B. Schroeder, E. Esarey, W. P. Leemans, Phys. Rev **A 87**, 062110 (2013)

7. L. Landau, E.M. Lifshitz, The classical theory of fields, (Pergamon Press, Oxford, 1975) p.227

8. E. Esarey, S.K. Ride, P. Sprangle, Phys. Rev. E 48, 3003 (1993)

9. P. Panek, J. Z. Kami´nski, F. Ehlotzky, Laser Physics 13, 457 (2003)

10. S. V. Bulanov, T. Zh. Esirkepov, M. Kando, J. K. Koga, S. S. Bulanov, Phys. Rev. E **84**, 055605 (2011); A. R. Bell, J. G. Kirk, Phys. Rev. Lett. **101**, 200403 (2008); J. G. Kirk, A. R. Bell, I. Arka, Plasma Phys. and Contr. Fusion **51** 085008 (2009); R.

11. F.V. Hartemann, A.K. Kerman, Phys. Rev. Lett. 76, 624–627 (1996); G.A. Krafft, Phys. Rev. Lett. 92, 204802 (2004); A. G. R. Thomas, C. P. Ridgers, S. S. Bulanov, B. J. Griffin, S. P. D. Mangles, Phys. Rev **X2**, 041004 (2012)

12. W.H. Press, S.A. Teukilsky, W. T. Vettering, B.P. Flannery, Numerical Recipes in Fortran, 2$^{nd}$ Ed. (University Press, Cambridge,1992) p.704

13. A. Yariv, Optical Electronics (Holt, Rinehart and Winston, NY,1985)

14. T. Tajima, J.M. Dowson, Phys. Rev. Lett.**43**, 267 (1979)

15. G. V. Stupakov, M. S. Zolotorev, Phys. Rev. Lett. 86,5274 (2001); P. X. Wang, Y. K. Ho, X. Q. Yuan, et al., Appl. Phys. Lett. 78, 2253(2001)

**Figure caption**

**Fig. 1** Spectra of backward scattering ($\theta=\pi$) at $a_0=1$ for Gaussian pulses 20, 90, and 200 fs; head-on collision at $\gamma=200$. (I) are the fundamental harmonics, (II)- the second harmonics. Radiation damping is included.

**Fig.2** Spectra of backward scattering ($\theta=\pi$) with $a_0=200$ Gaussian pulses of 60 fs (a) and 150 fs (b) for a head-on collision, $\gamma=700$, with and without the radiation damping. Inserts show the pulse field, $a_0(t)$, seen by the electron during the interaction and longitudinal momentum, $P_z$, of the electron.

**Fig. 3** Time evolution of $\chi$ parameter for an electron $\gamma=2000$ with and without the radiation damping in a 20 fs Gaussian pulse with $a_0=100$; head-on collision.

**Fig. 4** Spectra of backward scattering ($\theta=\pi$) with $a_0=100$ Gaussian pulses of 10 fs (a) and 40 fs (b) in a head-on collision, $\gamma=2000$, with the radiation damping, and (c) 10 fs laser pulse with $a_0=500$.

**Fig. 5** Trajectories and backward spectra (insert) for $a_0=100$ and $\tau=200$ fs in a head-on (1) collision and a collision with a 2 μm transverse shift.

**Fig. 6** Spectra of backward scattering ($\theta=\pi$) with $a_0=100$ (a) and $a_0=200$ (b) Gaussian pulses of 20 fs at a head-on collision with $\gamma=16000$ ($\varepsilon\sim 8$ GeV) with radiation damping.

**Fig. 7** Evolution of the electron longitudinal momentum in the two pulse scheme: both pulses have $a_0=100$, $\tau=10$ fs, and focused in $w_0=5$ μm, head-on collision; (1),(2),(3) are given for different delays between the pulses.

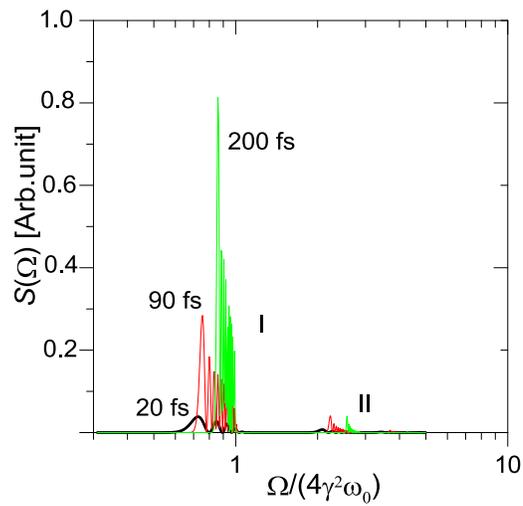

**Fig.1**

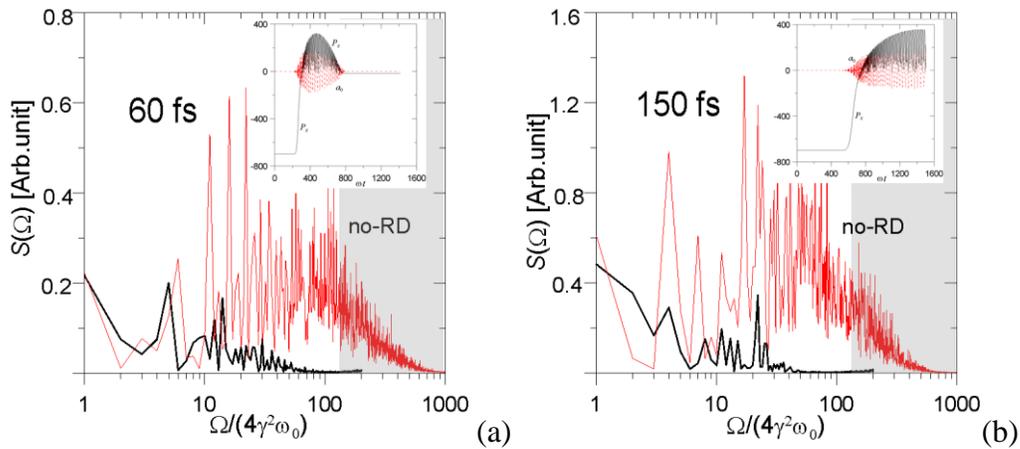

**Fig.2**

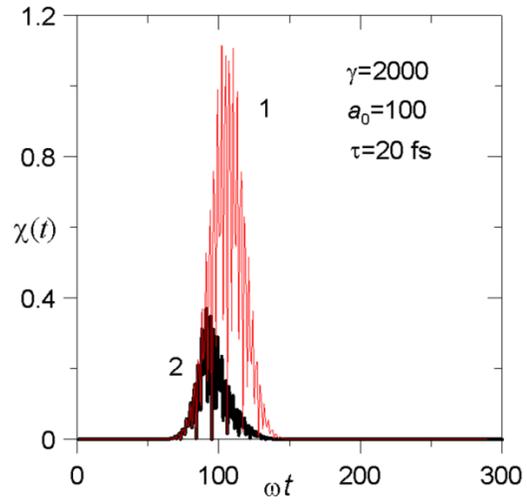

Fig.3

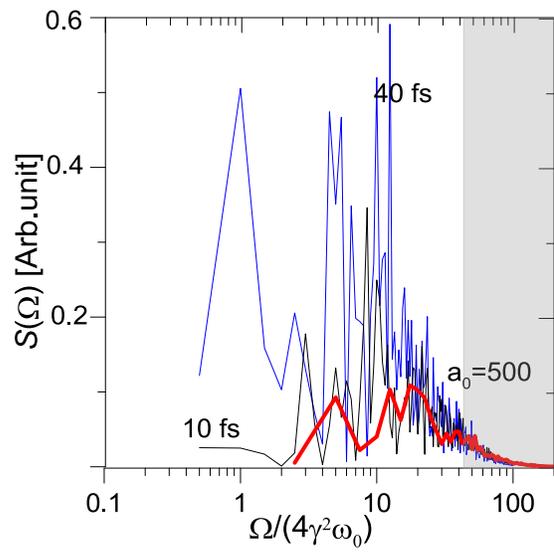

Fig.4

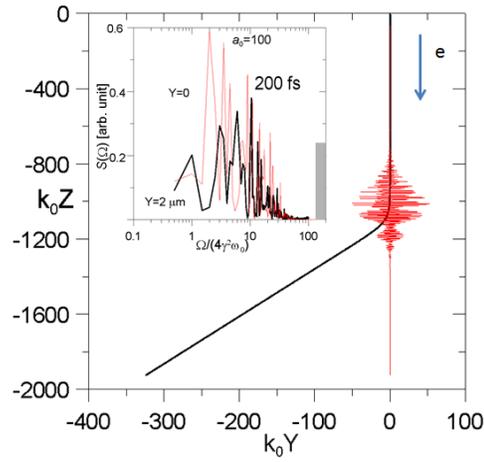

**Fig.5**

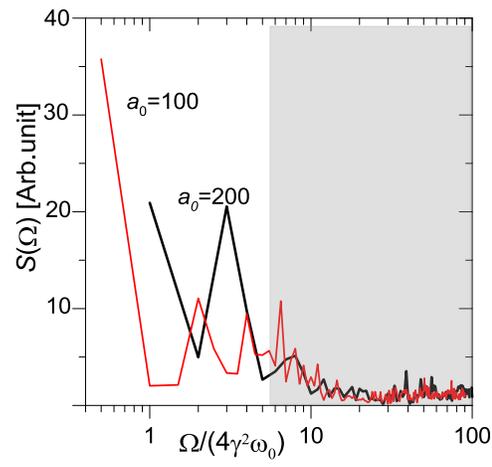

**Fig.6**

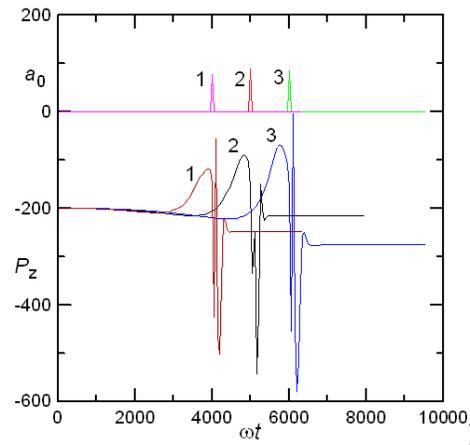

**Fig.7**